\documentclass[conference]{IEEEtran}
\IEEEoverridecommandlockouts
\usepackage{cite}
\usepackage{amsmath,amssymb,amsfonts}
\usepackage{algorithmic}
\usepackage{graphicx}
\usepackage{textcomp}
\usepackage{xcolor}
\usepackage{url}
\usepackage[skip=1ex,labelfont=bf, textfont=it]{caption}
\usepackage[textfont=rm]{subcaption}

\def\BibTeX{{\rm B\kern-.05em{\sc i\kern-.025em b}\kern-.08em
    T\kern-.1667em\lower.7ex\hbox{E}\kern-.125emX}}
\begin{document}

\title{CoMMA Protocol: Towards Complete Mitigation of \\
Maximal Extractable Value (MEV) Attacks\\
}

\author{\IEEEauthorblockN{Dev Churiwala}
\IEEEauthorblockA{\textit{Viterbi School of Engineering*}\thanks{*Visiting Student from BITS Pilani, Goa Campus.}  \\
\textit{University of Southern California}\\
Los Angeles, USA \\
f20180602@goa.bits-pilani.ac.in}
\and
\IEEEauthorblockN{Bhaskar Krishnamachari}
\IEEEauthorblockA{\textit{Viterbi School of Engineering} \\
\textit{University of Southern California}\\
Los Angeles, USA \\
bkrishna@usc.edu}
}

\maketitle

\begin{abstract}
MEV attacks have been an omnipresent evil in the blockchain world, an implicit tax that uninformed users pay for using the service. The problem arises from the miners' ability to reorder and insert arbitrary transactions in the blocks they mine. This paper proposes a 2-phased transaction protocol to eliminate MEV attacks. The user requests an interaction token from the on-chain counter-party. This token serves as a blind preemption for the counter-party and prevents the reordering of transactions at lower levels in the blockchain framework. We prove the correctness of the CoMMA protocol and demonstrate its efficacy against MEV attacks.
\end{abstract}


\section{Introduction}
With the explosion in the adoption and use of cryptocurrencies, the underlying blockchain infrastructures handle copious amounts of transactions. In traditional distributed consensus algorithms, `miners' insert these transactions into the blockchain ledgers. The miners pick unconfirmed transactions from the mempool, generally monetarily motivated, based on the profit they can make on the particular transaction. That said, nothing restricts the miners from including their transactions in the proposed block or from reordering the transactions in a way that is financially lucrative for them. The miner's economic incentive and uninhibited power over the included transactions and order give rise to MEV attacks~\cite{daian2020flash}. 

Some blockchain actors look to the opportunity for financial gain. They create MEV bots, automated scripts that specifically scout for target transactions that can be exploited to make money at the users' expense. Over the past month, in terms of USD volumes, these MEV bots have attributed for 46.1\% of all transactions running through Uniswap V3, the world's largest decentralized exchange (DEX)~\cite{adams2021uniswap}. This ratio translates to nearly \$20 billion in transactions from MEV bot activities alone~\cite{sui_2022}!

Traditionally, MEV activities have been overlooked because they are believed to stabilize the on-chain markets. However, drawing on free-market economics, eliminating these will not disrupt the market - other stabilizing factors are likely to bring about equilibrium~\cite{10.2307/3502458}. Our work analyses MEV activities and proposes an efficient solution to mitigate them.

\section{Related Works}
There has been much interest in mitigating MEV attacks recently. Zhou \emph{et al.} proposed A$^2$MM, which only targets insertion attacks~\cite{zhou2021a2mm}. Varun \emph{et al.} suggest using machine learning to detect and mitigate MEV attacks~\cite{varun2022mitigating}. Pillai proposes a new incentive structure for mitigating MEV activities~\cite{Pillai2022}. Malkhi and Szalachowski propose an augmented BFT protocol~\cite{malkhi2022maximal}. Furthermore, drawing on traditional cryptography, one may even consider encrypting threshold values to prevent identification of target transactions~\cite{shoup1998securing}. 

However, these techniques are either heavily computation-intensive or require the modification of low-level blockchain infrastructure. Our work seeks to eliminate MEV attacks efficiently, and it incorporates right into the existing blockchain framework.

\section{Understanding MEV Attacks}
\subsection{Categories of MEV bots}
\begin{itemize}
    \item \textit{Miner-controlled}: The miner deploys a bot to scout the mempool for candidate transactions that can be exploited. Upon finding such a transaction, the bot includes it in the proposed block with one or more malicious transactions inserted before or after the target transaction. The miner proposes the block in the order determined by the bot. If accepted, the block is incorporated into the ledger, and the miner makes a profit at the user's expense. 
    \item \textit{User-Controlled}: These bots continuously monitor the mempool for candidate transactions. Once a target transaction has been found, the bots submit one or more of their malicious transactions with just enough additional incentive for this transaction to be picked up first by a miner, thus making the bot's owner a profit. 
\end{itemize}

\subsection{Typical MEV attack on a DEX}
\begin{figure}[ht]
\centering
    \setkeys{Gin}{width=\linewidth}
\centerline{\includegraphics{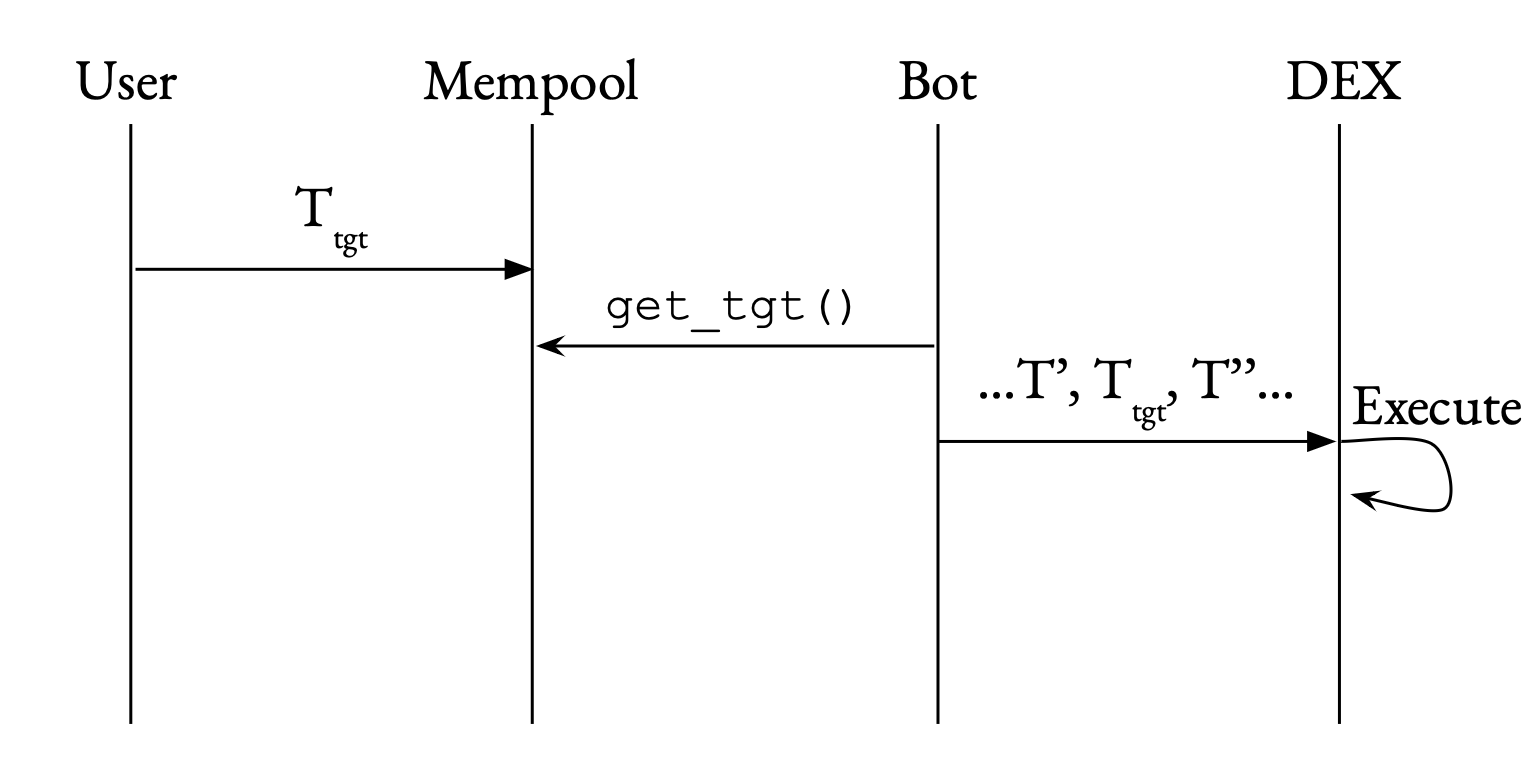}}
\caption{A transaction without the CoMMA protocol.}
\label{fig1}
\end{figure}

Fig.~\ref{fig1} depicts a miner-controlled bot exploiting a typical user. Consider a user who wishes to insert a transaction $T_{tgt}$ onto the ledger. The user submits transaction $T_{tgt}$ to the mempool, where it resides until a miner picks it up. The bot continuously monitors the mempool for `target' transactions, shown at function \texttt{get\_tgt()} in the figure. 

When the bot detects a potentially exploitable transaction (perhaps one whose slippage tolerance was too high), it picks the transaction (here $T_{tgt}$) for the following block proposal. The bot also adds transactions $T'$ and $T''$ to the block and orders the transactions $\{T', T_{tgt}, T''\}$ - `sandwiching' the target transaction between its transactions. When this block is accepted and run, this `sandwiched' transaction makes the miner money, and the MEV attack is successful~\cite{zust2021analyzing}. Similarly, it may suffice for the attacker to insert a transaction before $T_{tgt}$ to carry out the attack~\cite{eskandari2019sok}.

\section{Proposed Solution}
\subsection{CoMMA v1}
MEV attacks are bound to occur as long as the malicious actors can monitor and identify transactions in the mempool. To combat this, we propose a 2-stage transaction architecture. We use a DEX as the on-chain counter-party to elucidate the protocol, but an analogous protocol extends to all on-chain applications.
\begin{enumerate}
    \item The user first generates an `order request,' $Rq_{or}$, for the DEX. This request is meant to alert the DEX of an `intended' order.
    \item This request is then submitted to the mempool as a transaction.
    \item The transaction gets picked up by a miner, who directly runs it on the DEX because there is no money to be made here.
    \item In turn, the DEX issues an `order token,' $Tk_{or}$, to the user, indicating their position in the queue for orders.
    \item The user then submits their order, $T_{x}$, and it is executed when the user reaches the head of the queue. 
    \item The DEX issues tokens well in advance and does not wait for late transactions. Any transaction not at the head of the queue in their designated turn loses the spot and has to re-initiate an `order request,' $Rq_{or}$.
\end{enumerate}

\begin{figure}[ht]
\centering
    \setkeys{Gin}{width=0.85\linewidth}
\centerline{\includegraphics{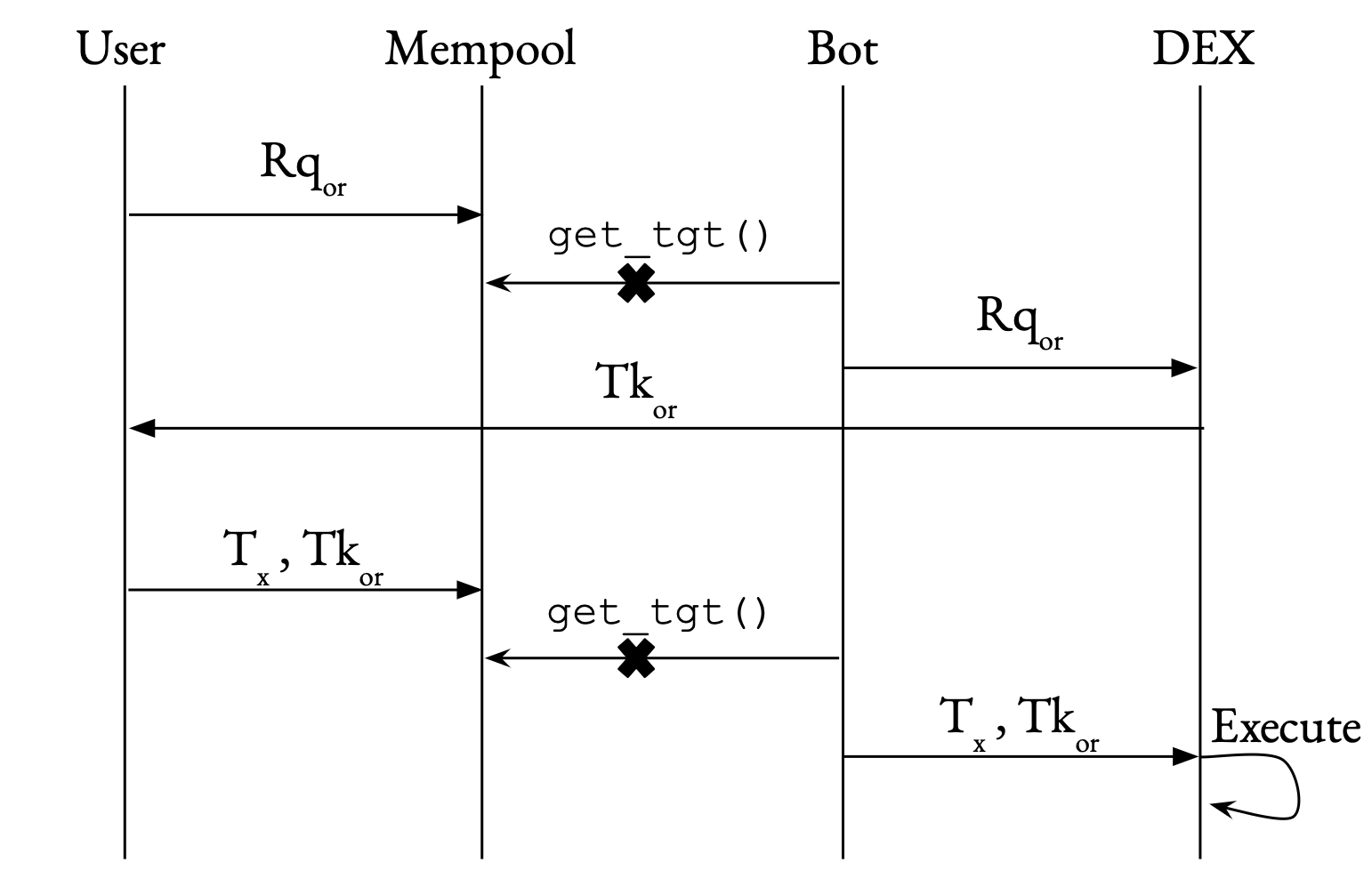}}
\caption{CoMMA v1}
\label{fig2}
\end{figure}

This protocol has the clear advantage of mitigating MEV attacks by determining an order queue. However, a glaring drawback of this protocol is the lack of verifiability. Given enough financial incentives, a malicious actor can still successfully carry out an MEV attack. Of course, this would cost the malicious actor since they need to know in advance whether the target transaction will be profitable. They still have to spend transaction fees issuing order requests to obtain queue slots for front-run target transactions that do not result in profit.

\subsection{CoMMA v2}
To overcome the limitation described above, we modify CoMMA v1. When making the initials `order request,' the user adds a one-way cryptographic hash of the intended order to the transaction, $Hx_{T_x}$. In doing so, the DEX can verify the hash before executing the order. This additional verification further discourages malicious MEV bots from attacking users. If the bots were to carry out an attack, they would have to guess the user's order details while creating the `order request' prior to seeing them in the next phase - with an infinitesimal chance of being right. These odds take away nearly all of the attacker's financial incentives for carrying out an MEV attack.

\begin{figure}[ht]
\centering
    \setkeys{Gin}{width=0.85\linewidth}
\centerline{\includegraphics{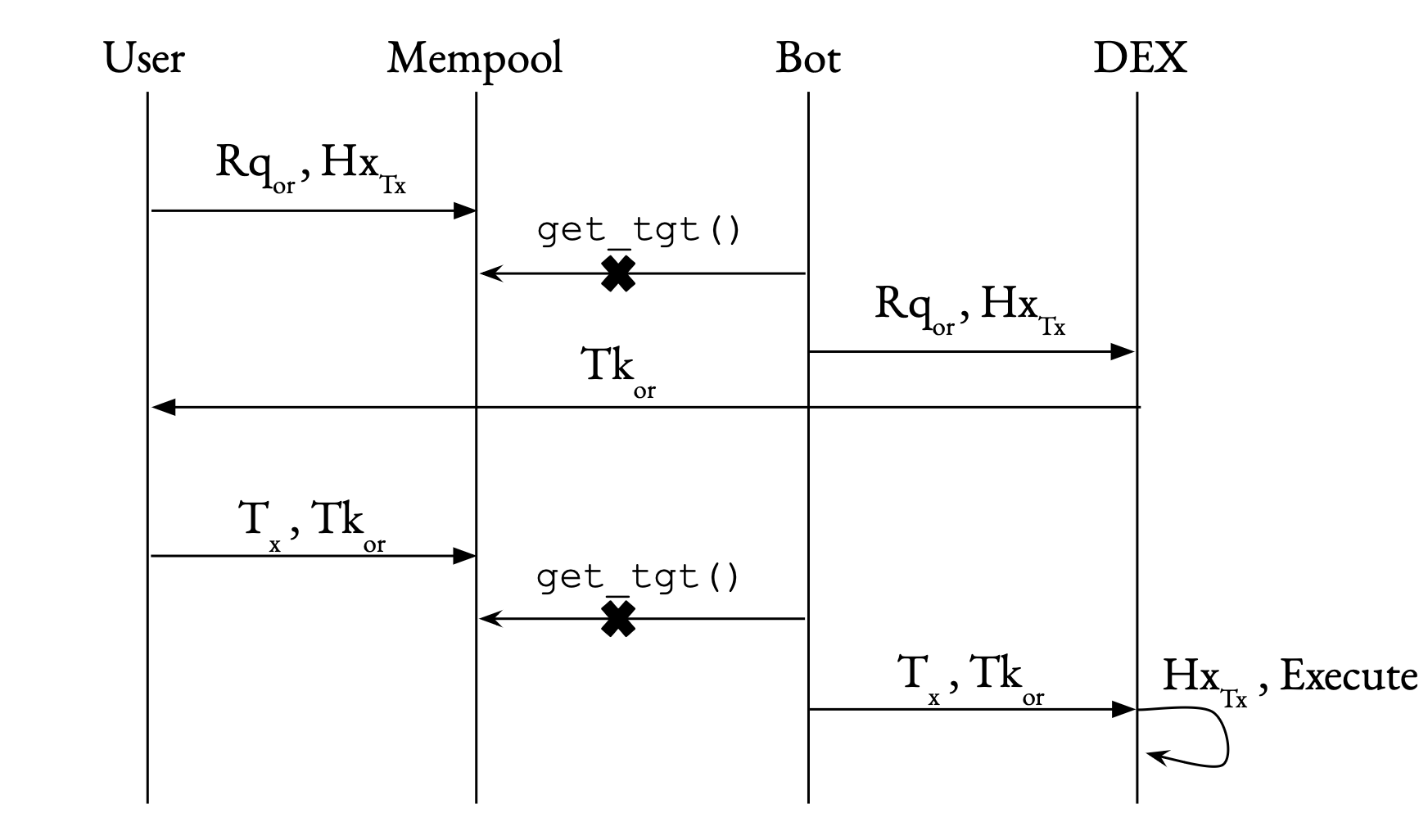}}
\caption{CoMMA v2}
\label{fig3}
\end{figure}

CoMMA v2's architecture adds additional latency and a minor pecuniary cost over traditional transaction mechanics. However, in exchange, it eliminates MEV attacks while upholding the foundational values of blockchains - trustlessness, decentralization, and transparency.

\section{Formal Verification}
\textit{Expectation}: We expect CoMMA v2 to eliminate MEV attacks while maintaining trustlessness, decentralization, and transparency.

\textit{Assumption}: The malicious actor, Oscar, is financially motivated.

\textit{Set-up}: Oscar should be able to insert formulated transactions before and after the target transaction.

\textit{Analysis}: We follow CoMMA v2 under the given assumption for the analysis.
\begin{itemize}
    \item \textit{Case (i).} The user, Alice, submits an order request to the mempool. At this point, Alice's request is visible to everyone monitoring the pool but only indicates an intent to trade, along with the hash of the intended trade. Oscar cannot use this information to formulate an MEV attack due to the irreversibility of the one-way hash function. If, at this point, Oscar were to obtain the transaction slots before and after Alice's transaction, he would have to guess the trade Alice would make to create his transaction hash values. Under the assumption that Oscar is financially driven, he would not choose to obtain these slots. The transaction is mined, and an order token is issued to Alice. The DEX stores Alice's Hash value with her token. Alice now submits her trade to the mempool, along with the token verifying her position and the trade. Since Oscar did not obtain the slots before and after Alice's, he cannot set up the MEV attack as described above. Alice carries out her trade without being exploited by Oscar. The transaction occurs without any direct interaction between the actors. Hence, \textit{case (i)} corroborates our expectation. \\

    \item \textit{Case (ii).} We also consider the alternate case, where Oscar decides to take a chance and obtain the adjacent slots. To do so, Oscar guesses Alice's trade and formulates his transactions accordingly. By the law of large numbers, we can safely state that Oscar will be incorrect nearly every time, and overall, he will lose money in transaction fees. Since we assumed that Oscar is financially motivated, and \textit{case (ii)} contradicts our assumption; we prove the correctness of CoMMA v2.
\end{itemize}

\section{Conclusion}
High liquidity flowing through blockchain systems create numerous MEV attack opportunities, often summing up to billions of dollars. Regular, uniformed users have to bear the entirety of this financial brunt. Although it remains to be empirically decided whether MEV attacks do more harm than good, we suggest that it is a vestigial vice. We propose CoMMA v2 to efficiently eliminate MEV attacks without modification to lower-level blockchain frameworks. Furthermore, we prove the correctness of CoMMA v2 for financially motivated market players and show that it upholds the foundational values of blockchains. We see CoMMA v2 as a modular upgrade for existing protocols and expect it to eliminate MEV attacks wherever incorporated. 

\bibliography{ref}
\bibliographystyle{ieeetr}

\end{document}